# Digital Musical Instrument Analysis: The Haptic Bowl

Gareth W. Young[1] and Dave Murphy[2]

University College Cork
Cork, Ireland.
[1] g.young@cs.ucc.ie
[2] d.murphy@cs.ucc.ie

**Abstract.** This experiment is a case study that applies a HCI-informed DMI Evaluation Framework. This framework applies existing HCI evaluation methods to the assessment of prototype Digital Musical Instruments (DMIs). The overall study will involve a three-part analysis: a description and categorisation of the device, a functionality evaluation that included an examination of usability and user experience, and finally an exploration of the device's effectiveness as a digital instrument. Here we present the findings of the first two parts of the framework, outlining the constituent components of the interface and testing the functionality of the device. The final stage of analysis will involve a longitudinal study, and will be carried out in order to assess the musical affordances of the device.

## 1 Introduction

This paper presents the results of a series of experiments to evaluate The Haptic Bowl, a prototype Digital Musical Instrument (DMI). To formally structure the experiment, we incorporated the framework of analysis as proposed in [1]. The main goal of the experiment presented here was to measure the device's DMI functionality and to assess the user's experiences of the device in a non-performance context. Information pertaining to usability and the user's experiences when using the device in a functional task was also captured. We also conducted post-task structured interviews to elicit important usability and user experience data. Our functionality testing included an analysis of usability and user experience in order to highlight any potential issues of function before the longitudinal study was carried out. According to the framework, to assess the effectiveness of the device as a DMI we must also undertake a longitudinal study of the device in a musical context. This stage must be undertaken to give our selected musical participants time to evaluate the device in a performance and compositional context rather than a sterile laboratory environment.

The first stage of our evaluation was to create a general description of the device's characteristics. Following this, our device was reduced down to its physical input variables. The next step was to contextualise our device for evaluation. This first stage of the analysis served to inform the evaluation and comparison methodologies that followed. Once informed, we evaluated our device's capability of undertaking potential HCI analysis tasks. A functionality experiment was devised to assess The Haptic Bowl's function as a DMI. This functionality testing served to quantify elements of the operational characteristics of the device. Following the functionality experiment, participants were invited to use The Haptic Bowl as a musical device. Post-study interviews were recorded and the participants then verbally evaluated the device.

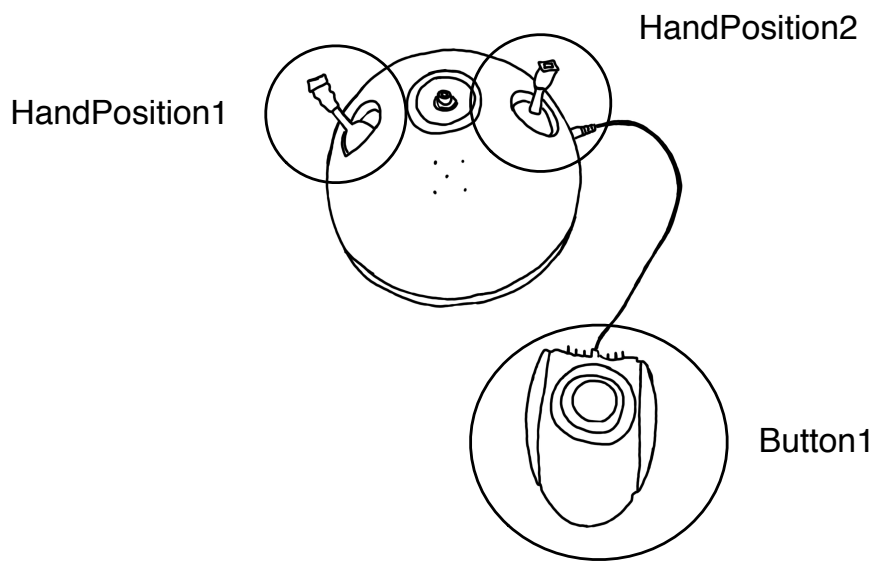

| | | |
|---|---|---|
| Type | HandPosition1 = HandPosition2 | Button1 |
| Manipulation | $dP_{xyz}$ | $P_z$ |
| Input | x, y [0 > 70 Deg], z [0 > 3 m] | z [0-Max$_z$] |
| State | d | up/down |
| Resolution fn | $f_{xyz}$ | T1($f_z$) |
| Output | x, y [ 14.6/Deg], z [341/m] | z [0-1] |
| Works | User defined | User defined |

**Fig. 1.** Analysis of primitive input for Haptic Bowl, two x, y, z position capturing devices and a one-dimensional foot switch.

## 2 Device Description

The first stage of our assessment was to formulate a general categorisation of the device in order to identify The Haptic Bowl's basic input elements. Following this, the input characteristics of the DMI were reduced to the physical variables to be manipulated. Finally, the taxonomy of the device was used to further reduce these classification variables down to their basic forms. At this point, we were then prepared to select an appropriate HCI method of functionality analysis that best fit our device.

### 2.1 Basic Description

The Haptic Bowl is an isotonic, zero-order, alternative controller that was developed from a console game interface. The internal mechanisms of a GameTrak tethered spatial position controller were removed and relocated into a more robust and aesthetically pleasing shell. All of the original HID circuitry was removed and replaced with an Arduino (Uno smd edition). This upgrade reduced internal latency, and allowed for increased device functionality. The controller has very little in the way of performer movement restrictions, as physical contact with the device is reduced to the two tethers that connect the user via gloves. Control of the device requires the performer to visualise an area in three dimensions, with each hand tethered to the device within this space. In addition to user proprioceptive awareness, haptic feedback components were incorporated to communicate musical performance data to the user. These included a strengthened spring force return mechanism for both tethers and audio frequency vibrotactile feedback delivered via actuators embedded in the gloves.

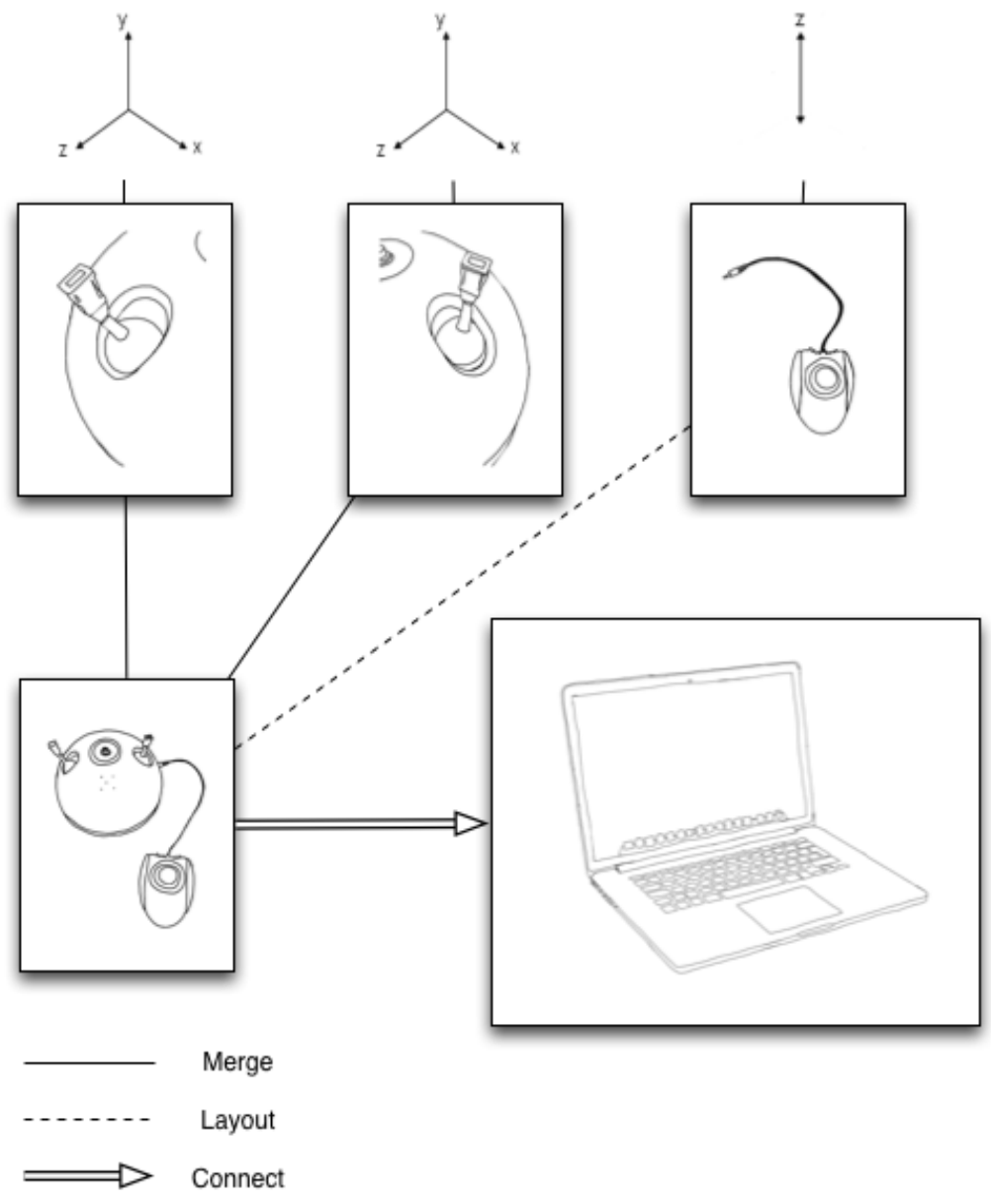

**Fig. 2.** Composition operators used to describe the Haptic Bowl

### 2.1 Primitive Movement Vocabulary

Figure 1 describes the basic controls incorporated into the Haptic Bowl using primitive movement vocabulary [2]. This vocabulary model includes six points of interest when discussing how an input device transduces physical movements into logical movement coefficients for measurement. As each hand is manoeuvred in a three-dimensional virtual space, the Cartesian X, Y, and Z coordinates of hand movements are measured. The manipulation operators of these planes of movement are measured continuously through the input domain operators applied for relative gesture input. Specifically, we can measure azimuth, medial, and frontal planes of movement over a domain set of 0° to 70° and a Z plane over a distance range of 0 to 3 meters. The 10-bit resolution of the analogue inputs on the Arduino (used to process input and interface with a PC) operates between a range of 0 - 1023. This range is repeated over the input domain range of each hand. The button mechanism operates over a single plane, in an up-down manner with a binary output of 0 or 1 (ON or OFF). The purpose of this device is user defined, which enables users to map multiple inputs to any application.

### 2.2 Composite Design

So far we have focused our discussion on the simple control and multidimensional aspects of the device. More specifically, the prototype is composed of a collection of one-dimensional elements that combine to form a multidimensional input device. For the analysis we consider three main compositional operator factors [3]:

1. Operators that can be connected.
2. Operators that can be laid out together.
3. The operator domain sets that can be merged.

The device is based on a combination of these three operators. Therefore, the Haptic Bowl involves the merging of generic three-dimensional sliders. For an accurate description of our device, we are required to consider the composition operators as seen in Figure 2.

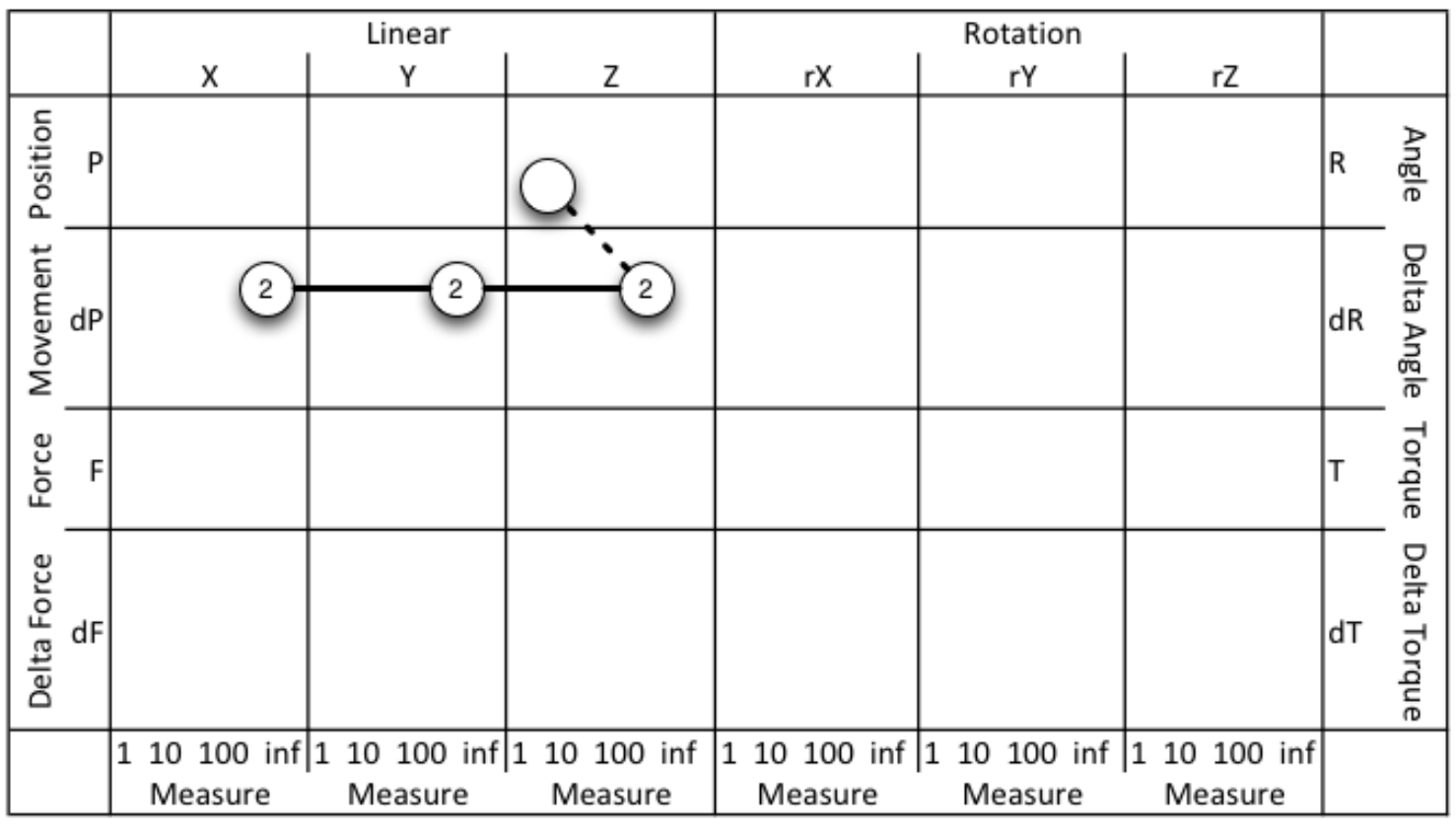

**Fig. 3.** Input device taxonomy used to describe the Haptic Bowl

### 2.4 Input Device Taxonomy

Figure 3 presents the input device taxonomy of The Haptic Bowl, derived from the analysis techniques discussed earlier. The device's input properties are presented on the vertical axis as physical characteristics and along the horizontal as linear or rotary measure [3]. The circles indicate that the device is capable of gesture capture in three dimensions. We have placed circles to highlight the linear movement along the X, Y, and Z planes. Additionally, the horizontal placement of the circles indicates the number of values that can be measured (0 > 1023). The path between the discrete elements indicates the manner in which they are connected. Black lines represent a merge composition of the X, Y, and Z components. Circles represent the individual elements with the numbers within signifying the number of identical inputs. The dashed line represents the layout composition of the foot-switch, whose values range from 0 - 1.

## 3 Functionality Testing

In order to assess the functionality of the Haptic Bowl, an experiment was devised that required participants to use the interface in a task. This was chosen to provide quantitative data that could be accurately used to compare to other devices. From analyzing the functional mechanisms of the device, a Fitts Law style experiment was designed. Additional to this, validated usability and user experience questionnaires were completed post-task that contained the following questioning techniques:

- Single Ease Question [4].
- Subjective Mental Effort Question [4].
- NASA task load index Questionnaire (pen and paper) [5].
- User Experience Questionnaire [6].
- Open ended questioning and informal interview.

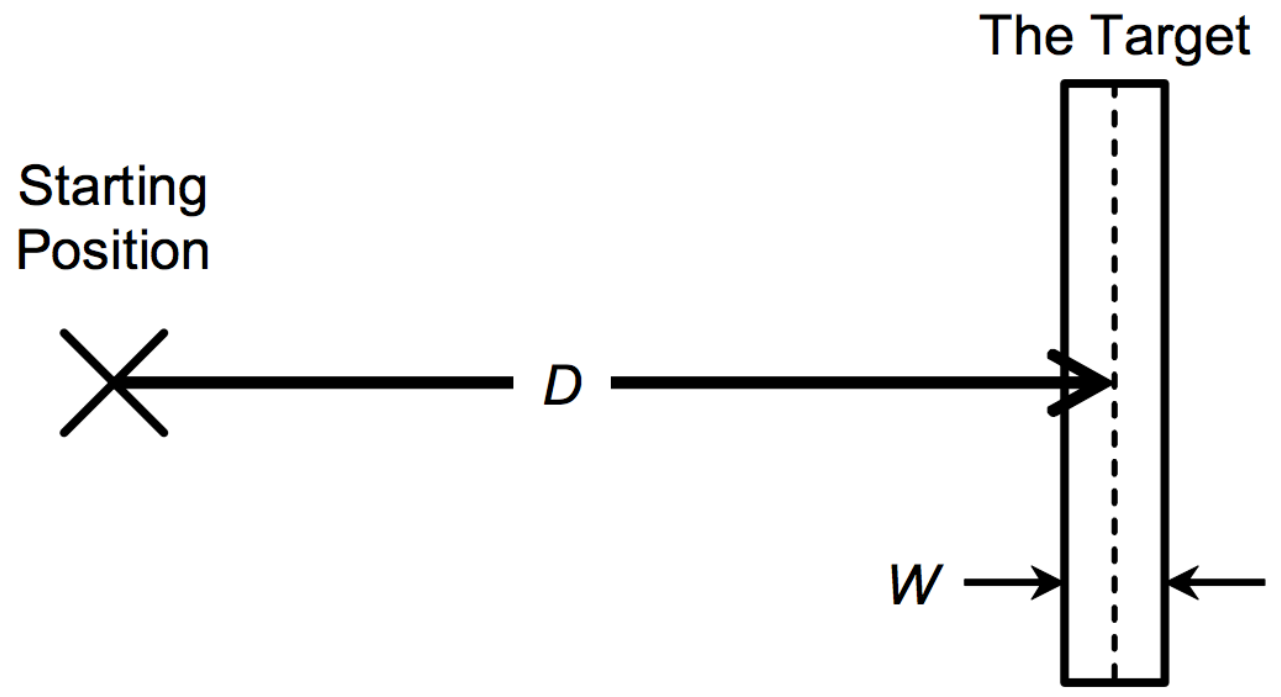

**Fig. 4.** Fitts' Law movement model

### 3.1 Incorporating Fitts' Law

Fitts' Law is used in HCI to describe the relationships between movement time, distance, and target size when performing rapid aimed movements (see figure 4). According to Fitts, the time it takes to move and point to a target of a specified width and distance is a logarithmic function of the spatial relative error [7]. Whilst this logarithmic relationship may not exist beyond a Windows, Icons, Menus, Pointer (WIMP) system, the experimental procedures can be followed to produce data for analysis. We measure a participant's ability to rapidly aim their movements within a predetermined audio spectrum towards a specified target frequency. Essentially, we have remapped physical distance to audio frequency distance, where the start position corresponds to below 20 Hz and a target position that lays slightly less than 1 kHz. The targets' width is predetermined as a physiological constant of 3 Hz for sine-wave signals below 500 Hz, which increases by approximately 0.6% (about 10 cents) as frequency increases towards 1 kHz [8].

### 3.2 Participants

Ten musicians participated in this experiment. All subjects were recruited from University College Cork and the surrounding community area. All of the participants classified themselves as musicians, having been formally trained or regularly performing in the past five years. Participants were aged 22 to 36 ($M$ = 26.3, $SD$ = 4.31). The group consisted of 8 males and 2 females.

### 3.3 Experimental Procedure

All stages of the experiment were conducted in an acoustically treated studio. The USB output from the Haptic Bowl was connected to a 2012 MacBook Pro Retina. The input data from the Arduino was converted into Open Sound Control (OSC) messages and outputted as User Datagram Protocol (UDP) information over port 12001. Pure Data (PD) then received this data. Within PD, both input values of Z-plane movement were used to create a virtual Theremin. The right hand controlled the pitch, and the left hand the volume. The operational range of the Haptic Bowl was normalised to fit within an operational range of 30 cm, which lay slightly above an average waist height. The foot-switch was employed to indicate the start and end of each test.

After a brief demonstration, participants were given five minutes to familiarise themselves with the operation of the device. Following this, subjects were then given a further five minutes to practice the experimental procedure. The experiment required the user to listen to a specific pitch and then seek and select this pitch with the Haptic Bowl as quickly and as accurately as possible. The start position for all steps was with hands resting in a neutral position below the waist. Subjects used the foot-switch to start and finish the test for each pitch. For each run of the experiment, eleven pitches were selected in counterbalanced order across a range of 110 Hz (A2) to 987.77 Hz (B6). Participants performed four runs of 11 pitch exercises, with a brief rest between each run. A Processing patch was used to capture input movement data and the time taken to perform the task, this was then outputted as a .csv file for analysis. After the experiment, participants were asked to complete a post-task evaluation questionnaire and informal interview.

## 4 Functionality Results

The results from the Fitts' Law style evaluation can be seen in figure 5 and table 1. The average Move Time (MT) for the Haptic Bowl was 5760 ms across the predefined frequency range outlined earlier. On average, all participants were able to target and select frequencies within the predetermined target size of 3 Hz for all frequencies below 261.6 Hz (C4), with an average standard deviation of 1.7 Hz in this range. The accuracy of pitch selection decreased with pitch increments from here on. For 261.6 Hz (C4) up to 523.25 Hz (C5), the standard deviation in pitch selection increased to 3.91 Hz. Beyond this range, from 523.25 Hz (C5) to 975.83 Hz (B6), the standard deviation increased to an average of 9.71 Hz.

**Table 1.** Fitts' Law style experiment results.

| Target Frequency (Hz / Pitch) | Actual Mean Frequency (Hz) | Standard Deviation (Hz) | Move Time (ms) | Standard Deviation (ms) |
|---|---|---|---|---|
| 110 (A2) | 110.06 | 1.64 | 6333.50 | 973.09 |
| 130.81 (C3) | 129.61 | 2.69 | 6669.00 | 2420.03 |
| 174.61 (F3) | 174.39 | 1.22 | 5982.17 | 2259.59 |
| 220 (A3) | 220.83 | 1.27 | 5478.00 | 2166.27 |
| 261.6 (C4) | 262.22 | 4.10 | 5162.78 | 1554.78 |
| 349.23 (F4) | 348.39 | 3.78 | 5206.83 | 1557.50 |
| 440 (A4) | 438.28 | 3.84 | 5244.61 | 1701.58 |
| 523.25 (C5) | 518.89 | 8.29 | 6032.22 | 1934.24 |
| 698.47 (F5) | 691.39 | 11.72 | 6058.56 | 1233.63 |
| 880 (A5) | 879.89 | 3.58 | 5408.89 | 1689.12 |
| 987.77 (B6) | 975.83 | 15.23 | 5786.72 | 1662.97 |

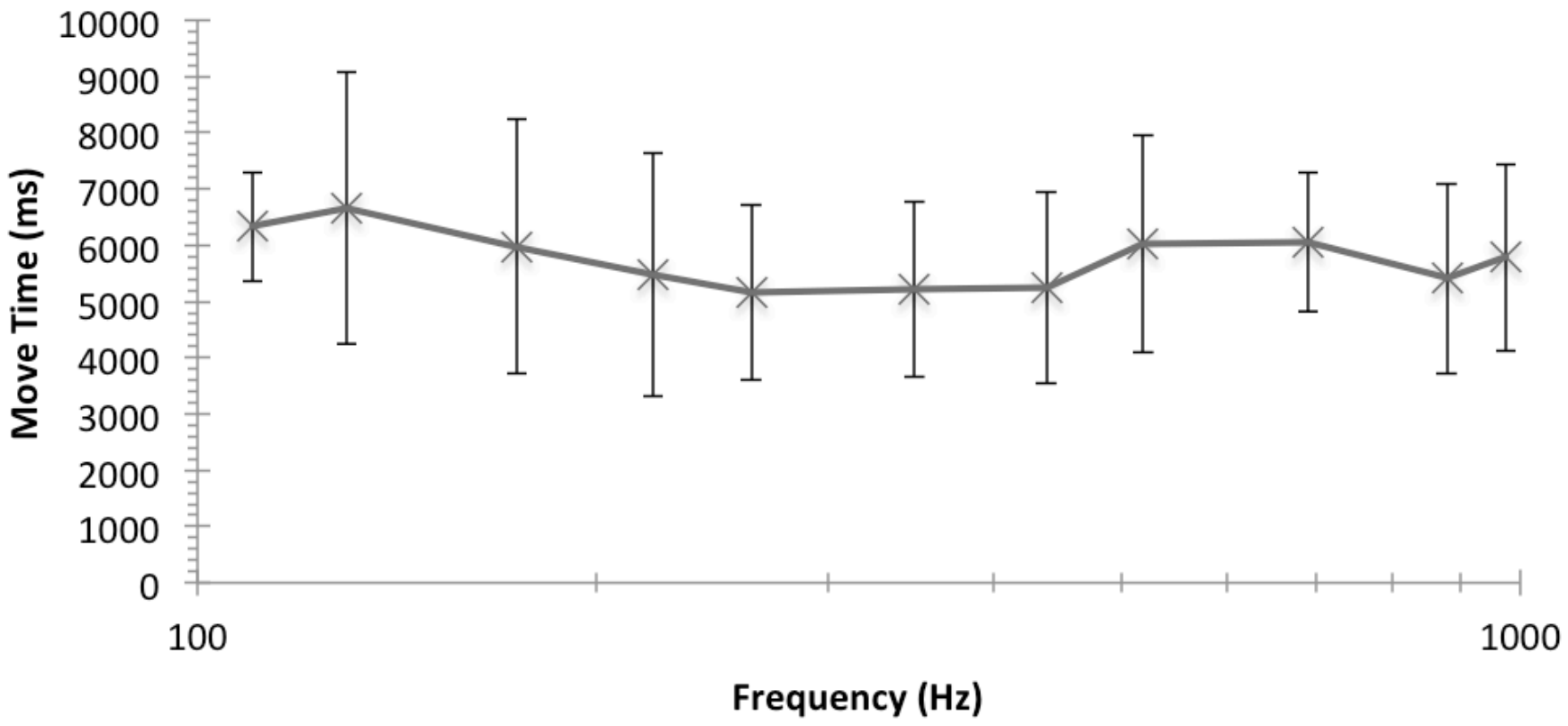

**Fig. 5.** Mean Move Time (MT) over frequency.

### 4.1 Usability Results

Post-task questioning concerning usage preference and the SEQ can be seen in figure 6. The majority of users were pleased with the device's performance and thought that they would use the device *somewhat* or *most often* (50%). 20% of the users did not think that they would use the device *neither often* nor *not very often*. These users indicated that they do not have an opinion about usage preference, as they would not normally use a computer interface to make music. Finally, 40% of users indicated that they would only use the interface *somewhat seldom*. These users were not particularly inspired by the functionality experiment, but indicated that if they could expand or explore the device's parameters further they might have rated it more favourably.

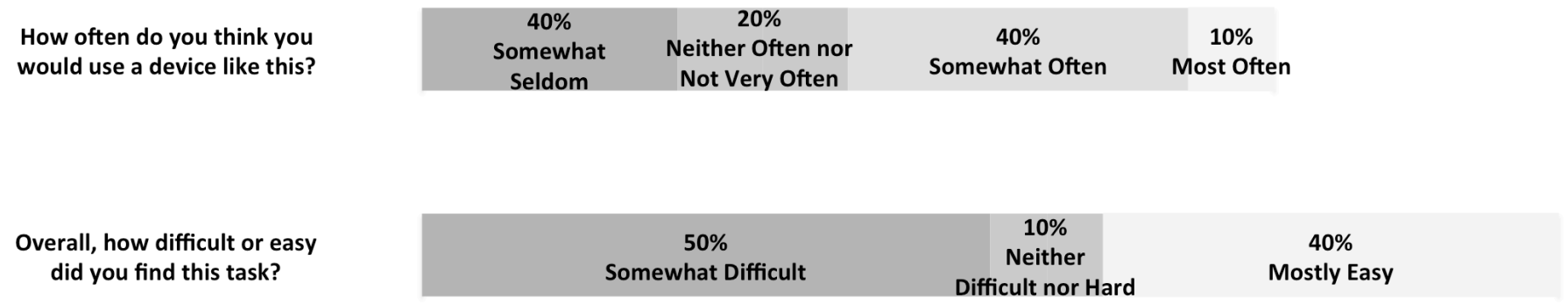

**Fig. 6.** Diverging stacked bar chart for estimated usage and the SEQ.

With the SEQ, the users were given the opportunity to consider their own performance and factored this into their responses. Users had to adjust their own rating of performance based upon the range of answers available (7 in total) and respond to their interpretation of the difficulty of the task accordingly [4]. The majority of users (50%) found that the task was *somewhat difficult* for them to complete. Users expressed that they were fully engaged in the task, as they would be if they were performing for the first time with any instrument. This increase in cognitive load moved them to consider their performance more critically, regardless of individuals' overall test results. 10% of the users consider the task to be *neither difficult* nor *easy*. Finally, 40% of users were pleased with their performance and rated the task as *mostly easy*.

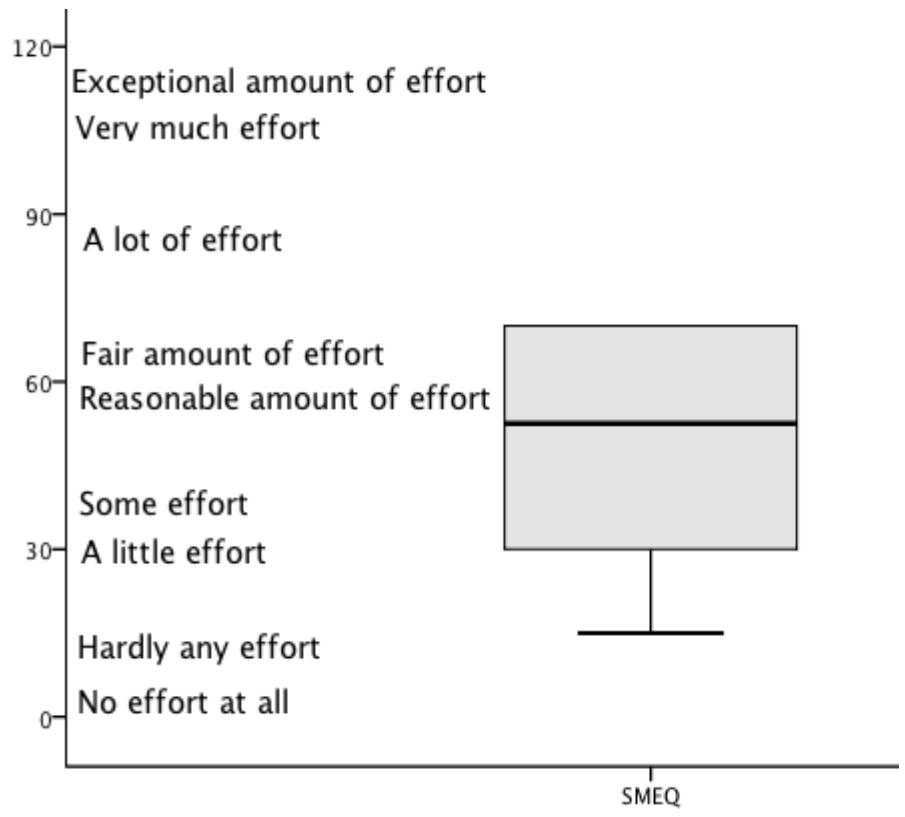

**Fig. 7.** Boxplot representing mean SMEQ answers.

The SMEQ presents a near-continuous response choice to the user, as seen in figure 7. Theoretically, this allowed the user to be more precise with their sentiments towards the device's usability. The premise of this scale was to elicit an indication of the user's thoughts towards the amount of mental effort they exerted during the task. The mode value of the SMEQ was calculated as 70, which corresponds to an evaluation of *fair amount of effort* for the majority of users (similar to the SEQ findings). However, this scale measures the amount of effort users felt they invested rather than the amount of effort the task demanded from them.

Following the self-evaluated measure of mental effort, we proceeded to measure the user's perceived subjective workload with the paper and pencil NASA Task Load Index (NASA-TLX) assessment tool. The NASA-TLX is designed to assess the device's in-task effectiveness via post-task questioning. The total workload, divided into the six TLX subscales, can be seen in figure 8.

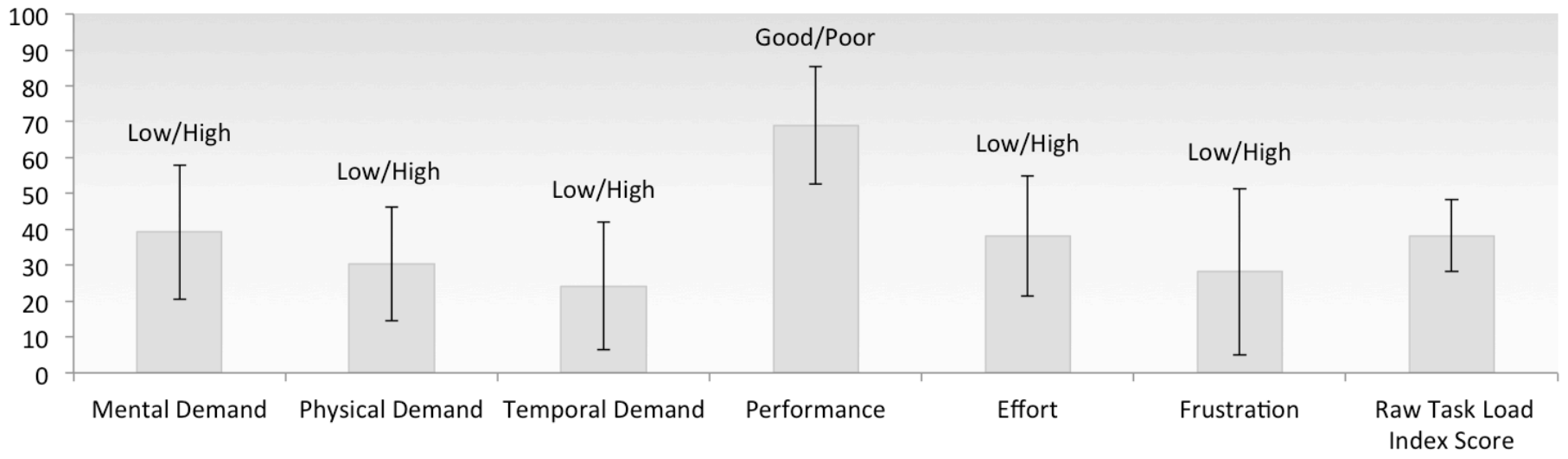

**Fig. 8.** Mean NASA-TLX subscale ratings for usability

The first indicator in the NASA-TLX subscale required the user to signify the level of demand the task required of them in terms of complexity. The mean level of perceived mental activity required for the task was measured as being 39.2%. Next, the mean physical demand of the task was measured as being 30.4%. Our users found the task relatively easy on a scale that measured how demanding the task was to physically complete. In terms of temporal demand, the time pressure felt performing the task; the mean user rating of the pace of the experiment was 24.2%. This score indicates that our users were not rushed and had plenty of time to complete the task without any pressure. In the TLX questionnaire, 69% of users indicated that they were relatively unsatisfied with their performance. The user's satisfaction with their performance corroborates with our earlier findings of negative user self-performance evaluations in the task's difficulty and mental effort. In contrast to this, a mean of 38.1% of the users indicated that they worked only somewhat hard mentally and physically to accomplish their level of performance. Additionally, our users indicated that they were not irritated or stressed by the task. In fact, a mean of 28.2% was recorded for frustration levels, weighting towards a relaxed attitude during the experiment. Finally, a mean overall 'raw TLX' measure of workload was calculated as 38.18%. A pairwise comparison of each dimension was not deemed necessary and thus, not completed.

### 4.2 User Experience Results

The final stage of the functionality analysis for this device incorporated an end-user assessment of the user's experiences during the task. A questionnaire was used to measure user experience quickly, simply, and as immediately as possible. Six important aspects of experience were captured, these included Attractiveness, Perspicuity, Efficiency, Dependability, Stimulation, and Novelty. Table 2 shows the results of the user experience questioning and as a measure of scale reliability, we also give the Cronbach's alpha coefficient score per scale. This score generally concerns itself with alpha values over 0.6. As can be seen in table 2, the internal consistency of questioning is acceptable for Attractiveness, Perspicuity, and Stimulation. However, there are poor internal consistencies for Efficiency, Dependability, and Novelty. For user experience measures on this scale, mean values between -0.8 and 0.8 are representative of a neutral evaluation of the corresponding dimension. Values greater 0.8 represent a positive evaluation and values below -0.8 represent a negative evaluation. The range is measured as -3 (very bad) and +3 (very good).

**Table 2.** User Experience data.

| UEQ Scale | Mean Score | Cronbach's Alpha |
|---|---|---|
| Attractiveness | 1.650 | 0.86 |
| Perspicuity | 1.625 | 0.63 |
| Efficiency | 1.425 | 0.29 |
| Dependability | 1.025 | 0.34 |
| Stimulation | 1.550 | 0.80 |
| Novelty | 1.775 | 0.43 |

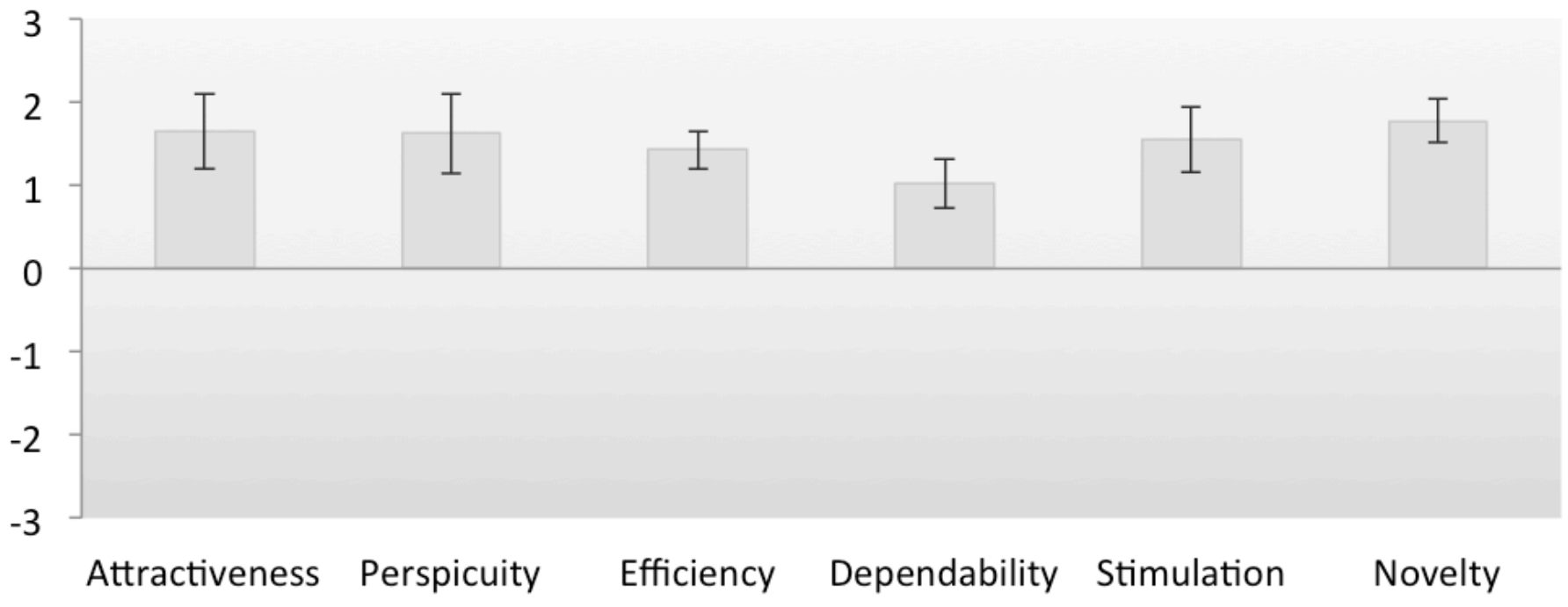

**Fig. 9.** User experience results.

### 4.2 Functionality Interviews

The final part of the functionality experiment was an interview-style evaluation of the Haptic Bowl (refer to figure 10). Participants were queried on their opinions expressed in open-ended questions presented to them in a questionnaire, which were then expanded in an interview. Users were inclined to be positive about the device (66.03%) and were pleased with the amount of feedback that was delivered to them via the haptic feedback.

*"You can feel the vibration produced, just like an instrument"*
*"Being able to hear and feel the sound makes for a pleasant playing experience"*

Other positive comments were directed towards how challenging the task was, but with positivity of how stimulating it was.

*“I found it challenging to play, but rewarding and motivating”*

22.44% of comments were neutral statements, such as comparisons to other devices that the user has used or their previous experiences with DMIs in general.

*“Most electronic instruments I’ve used don’t have tactile feedback”*
*“This is unlike any instrument I’ve used, but maybe a stylophone, i.e. a tethered control mechanism producing a similar sound”*

Only 11.54% of user comments were negative. These comments were mainly directed towards potential difficulties that the device may have in exercises that require jumps in pitch (frequency) that span a number of octaves instantly. Concerns about following specific score information, such as pitch and tempo, were also expressed.

*“Would be difficult to play fast leaps [of pitch]”*
*“It was sometimes difficult to control pitch”*
*“It would require training to incorporate tempo”*

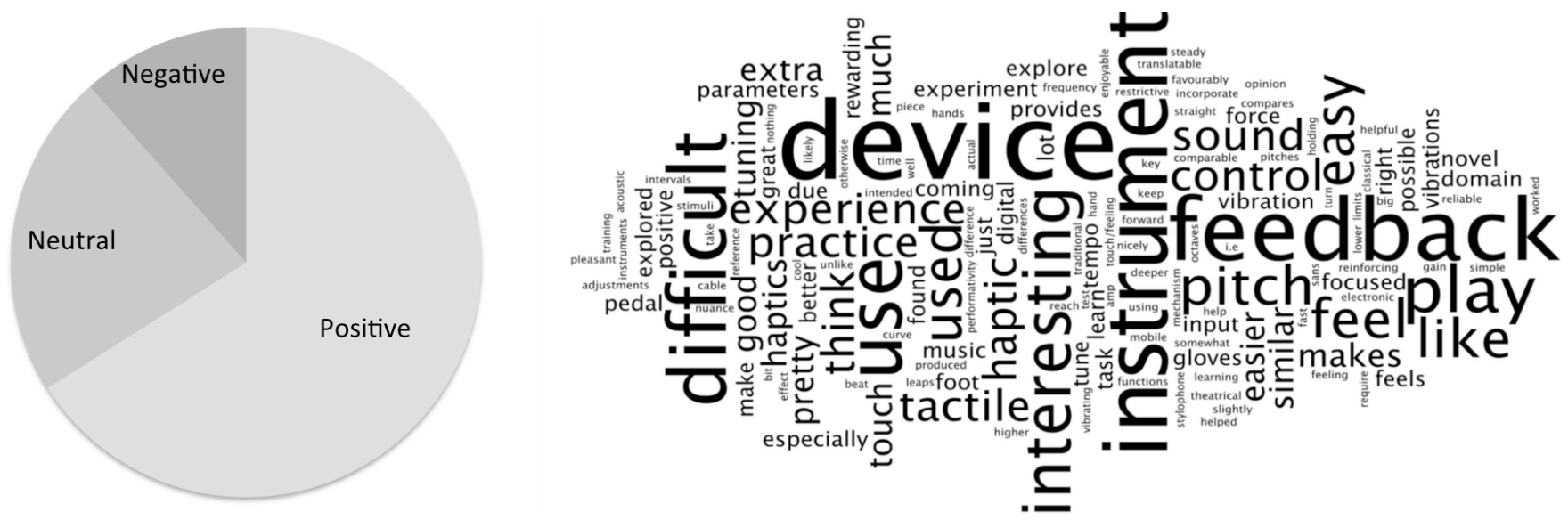

**Fig. 10.** Positive, neutral, and negative comments and word cloud of interview data.

## 5 Conclusions

From our analysis of the Haptic Bowl we have demonstrated how the HCI informed framework suggested in [1] can be applied to DMI evaluation. Specifically, we have shown how a device analysis may be informed and developed from techniques that are applied in HCI evaluations in general computing and computing for unique applications. With reference to the study presented here, the functional capacity of the Haptic Bowl to afford users with the ability to select specific pitches was quantified and evaluated. Participants were capable of selecting specific frequencies (pitches) within a mean move time of 5760 ms (SD = 476.69). The mean accuracy of pitch selection of all participants was 5.21 Hz, with standard deviation increasing with frequency (pitch). Post-task questioning of usage preference found that 50% of users were pleased with the device’s performance and would use the device often. For the SEQ, 50% of participants found that the task somewhat difficult, but were engaged in the task as they would be if they were performing for the first time with a new instrument. The SMEQ confirmed these findings with a mode rating of 70, indicating a fair amount of effort was required to perform the task. Further usability testing found a mean overall 'raw TLX' measure of workload of 38.18%. Consistently positive measures for Attractiveness, Perspicuity, and Stimulation were found in user experience questioning. With the specific categorisation of our device and the quantitative data taken during functionality testing, we are now afforded with accurate data measures that can be applied in future device comparisons.

Following this functionality analysis will be a longitudinal study of The Haptic Bowl. This will be carried out in order to assess the performance potential of the device in a musical context. As musicality is arguably subjective in its evaluation, we will not assess the user's application in composition, but the performance of the device in musical tasks. Additionally, the device's usability and the user's experience when using the device as a performance tool will be recorded. Participants will be given a preselected list of musical tasks to perform and then evaluate the devices performance in the execution of these tasks, as in [1].